\documentclass[a4paper]{article}

\newcommand{\lKG}{\ensuremath{\lambda_{{\rm KG}}}}
\newcommand{\lD}{\ensuremath{\lambda_{{\rm D}}}}

\newcommand{\be}{\begin{equation}}
\newcommand{\eea}{\end{eqnarray}}
\newcommand{\bea}{\begin{eqnarray}}
\newcommand{\ee}{\end{equation}}

\newcommand{\psiSch}{\ensuremath{\psi_{{\rm Sch}}}}
\newcommand{\psiKG}{\ensuremath{\psi_{{\rm KG}}}}
\newcommand{\psiD}{\ensuremath{\psi_{{\rm D}}}}

\newcommand{\e}[1]{\exp({#1})}
\newcommand{\D}{{\rm d}}

\newcommand{\AiryAi}{\ensuremath{\mathop{{\rm Ai}}}}

\newcommand{\oder}[2]{\frac{\D #1}{\D #2}}
\newcommand{\pder}[2]{\frac{\partial #1}{\partial #2}}
\newcommand{\dt}{\ensuremath{{\rm d}t}}
\newcommand{\dx}{\ensuremath{{\rm d}x}}
\newcommand{\dy}{\ensuremath{{\rm d}y}}
\newcommand{\dz}{\ensuremath{{\rm d}z}}
\newcommand{\phiD}{\ensuremath{\phi_{{\rm D}}}}
\newcommand{\ds}{\ensuremath{{\rm d}s}}

\usepackage{amssymb}
\usepackage{amsmath}
\author{M. Alimohammadi
\footnote{alimohmd@ut.ac.ir} and B. Vakili
\\  {\small Department of Physics, University of Tehran,}
\\ {\small North Karegar Ave., Tehran, Iran.}
} 
\title{Spin 0 and spin 1/2 particles in a constant scalar-curvature background
} 
\date{}
\begin{document}
\maketitle
\begin{abstract}
We study  the Klein-Gordon and Dirac equations in the presence of
a background metric $ds^2 = -dt^2 + dx^2 + e^{-2gx}(dy^2 + dz^2)$
in a semi-infinite lab ($x>0$). This metric has a constant scalar
curvature $R=6g^2$ and is produced by a perfect fluid with
equation of state $p=-\rho /3$. The eigenfunctions of spin-$0$ and
spin-$1/2$ particles are obtained exactly, and the quantized
energy eigenvalues are compared. It is shown that both of these
particles must have nonzero transverse momentum in this
background. We show that there is a minimum energy $E^2_{{\rm
min}}=m^2c^4 + g^2c^2\hbar^2$ for bosons $(E_{{\rm KG}} > E_{{\rm
min}})$, while the fermions have no specific ground state
$(E_{{\rm Dirac}} > mc^2)$.

\noindent

\end{abstract}
\section{Introduction}
Understanding the connection between the quantum mechanics and
gravity has been one of the main purposes of physics, from the
early birth of quantum mechanics, and many efforts have been made
in this area [1-6]. These investigations have a vast spectrum,
from the simplest case of the Schrodinger equation in the presence
of constant gravity \cite{1}, to the most complicated case of
studying the Berry phase of spin-$1/2$ particles moving in a
space-time with torsion \cite{7}. Although, in almost all cases,
the quantum gravitational effects are weak, but they can be
measured experimentally even for the weak gravity of our earth,
see for example \cite{8,9} for the earliest recent experiments.

One of the interesting and important questions which arises in
this connection is that how much the spin of the quantum particles
is important in the quantum-gravity phenomena. For example the
problem of equivalence principle have been studied for spin-$1/2$
particles in \cite{10}. It has been shown that there is some
difference between the Dirac Hamiltonian in a Schwarzschild
background and in a uniformly accelerating Minkowski frame, which
can be a signature of violating the equivalence principle for
spin-$1/2$ particles. This violation does not exist for spin-$0$
particles.

From another point of view, the spectrum of spin-$0$ and
spin-$1/2$ particles in a constant gravitational field has been
studied in \cite{11}, and it is shown that they differ by an
amount of $mg\hbar c$, where $m$ is the mass of particles
(fermions and bosons) and $g$ is the acceleration of gravity. This
is an important, although weak, difference which shows the
influence of the spin in gravitational interaction. In this paper,
we want to study this effect for a more complicated case.

The situation considered in \cite{11} is a semi-infinite
laboratory ($x>0$) with metric $\ds^2=u^2(x) (-\dt^2+\dx^2) +
\dy^2 +\dz^2$ and an infinite barrier in $x<0$ region. For a
constant gravity in a small region, $u (x) = 1+gx$, the
eigenfunctions have been obtained for the Klein-Gordon and Dirac
equation in a form of power series, and the eigenvalues have been
obtained in high energy and in vertically fall cases.

In this article, we are going to study the same semi-infinite lab,
but with a more complicated metric. We first consider the metric
$\ds^2=-\dt^2+\dx^2+u^2(x) (\dy^2+\dz^2)$, in which there are
coupling terms between $x-y$ and $x-z$ coordinates. To make it
simple, after some steps, we take an exponential form for $ u (x)
$, that is $u (x) =\exp (-g x) $. We show that this metric
corresponds to a perfect-fluid, with negative pressure, and has a
constant scalar-curvature $R=6g^2$. We solve the Klein-Gordon and
Dirac equations in this constant curvature background (in $x>0$
region) and find the eigenfunctions exactly. In some energy
regions, the energy eigenvalues of spin-$0$ and spin-$1/2$
particles are obtained and it is shown that they do not coincide.
One of the main features of spectrum is that the spin-$0$
particles have a ground state energy $E_{\rm min}^2=m^2 c^4 + g^2
c^2\hbar^2$, while the fermions' kinetic energy can be any
positive value, determined by their momentum. This is another
interesting signature which shows the importance of spin in
quantum-gravity phenomena.

The plan of the paper is as follows: In section $2$, the
Klein-Gordon (KG) and Dirac equations are obtained in
$\ds^2=-\dt^2+\dx^2+u^2(x) (\dy^2+\dz^2)$ background. For
simplifying the results, we take $u'(x)/u(x)$ as a constant, which
restricts us to $u (x) =\exp (-gx)$. In sections $3$ and $4$ we
calculate the eigenfunctions of KG and Dirac equations,
respectively, and in section $5$ the eigenvalues are compared.
Finally, to understand the physics behind this metric and to
justify the falling behavior of trajectories, we calculate the
energy-momentum tensor and equation of state of the matter
corresponds to this metric and also the geodesics of it.

\section {Relativistic quantum equations in arbitrary $u(x)$ background}
In a spacetime with metric $g_{\mu\nu}$, the Klein-Gordon equation
is
\begin{equation}
\left( \frac{1}{\sqrt{-g}} \pder{}{x^\mu}\sqrt{-g}\,
g^{\mu\nu}\pder{}{x^\nu} - m^2 \right)\psiKG = 0 ,\nonumber
\end {equation}
where $g := \det\left [g_{\mu\nu}\right]$ (in $c=\hbar=1$ unit).
The Dirac equation in a curved spacetime is
\begin {equation}
[\gamma^a \left(\partial_a + \Gamma_a\right)- m ]\psiD=0 ,
\nonumber
\end{equation}
in which the spin connections $\Gamma_a$ can be obtained from
tetrads $e^a$ through
\begin {align}
\D e^a + \Gamma^a{}_b \wedge e^b =& 0,\nonumber\\
\Gamma^a{}_b :=& \Gamma^a_{}{cb} \, e^c,\nonumber\\
\Gamma_a :=&-\frac{1}{8} \left [\gamma_b, \gamma_c \right]
\Gamma^c{}_a{}^b. \nonumber
\end{align}

We consider a gravitational field which is represented by the
metric
\begin{equation} \label{A}
\ds^2 = -\dt^2 + \dx^2 + u^2(x) \left( \dy^2 + \dz^2 \right).
\end{equation}
For this metric, the Klein-Gordon equation reads
\begin{equation} \label{B}
\left[-\pder{^2}{t^2} + \pder{^2}{x^2} + 2{u'\over
u}\pder{}{x}+\frac{1}{u^2} \left(\pder{^2}{y^2} + \pder{^2}{z^2}
\right) - m^2 \right] \psiKG = 0.
\end{equation}
To write the Dirac equation, we need the spin connections. For
metric (\ref{A}), the nonvanishing $\Gamma^a{}_b $ 's are
$\Gamma^2{}_1 = -\Gamma^1{}_2 = (u'/u) e^2 $ and $ \Gamma^3{}_1 =
-\Gamma^1{}_3 = (u'/u) e^3 $. Therefore $ \Gamma^2{}_2{}^1 =
-\Gamma^1{}_2{}^2 = \Gamma^3{}_3{}^1 = -\Gamma^1{}_3{}^3 = u'/u $,
from which $ \Gamma_2 = -(u'/2u) \gamma_1 \gamma_2 $ and $
\Gamma_3 = -(u'/2u) \gamma_1 \gamma_3 $. Noting that $\gamma^a
\partial_a =\gamma^a e^\mu{}_a \partial_\mu $, where for metric
(\ref{A}) we have $e^\mu{}_a ={\rm diag} (1,1,1/u,1/u) $, the
Dirac equation becomes:
\begin{equation} \label{C}
\left[\gamma^0 \pder{}{t} + \gamma^1 \pder{}{x} + \frac{u'}{u}
\gamma^1 + \frac{1}{u}\left(\gamma^2 \pder{}{y} + \gamma^3
\pder{}{z} \right) - m \right] \psiD=0.
\end{equation}

For a semi-infinite lab, which there is an infinite potential
barrier at $x=0$, we must impose the suitable boundary condition
at $x=0$. In nonrelativistic case, the Schrodinger equation, the
boundary condition is $ \lim_{x \to 0^+} \psiSch = 0$, which
emerges from the fact that the Schrodinger equation is second
order in $x$, so $\psiSch$ must be continuous at $x=0$. For
Klein-Gordon equation, which is also second order in $x$, the
suitable boundary condition is
\begin{equation} \label{D}
\psiKG(0) = 0.
\end{equation}
But the Dirac equation is of \textit{first} order and therefore
$\psiD$ can be discontinuous at $x=0$, if the potential goes to
infinity there. This problem has been investigated carefully in
\cite{11}, and it has been shown that the suitable boundary
condition for infinite potential barrier is
\begin{equation} \label{E}
 \left(\gamma^1 - 1 \right) \psiD(0) = 0.
\end{equation}
So our problem is to solve eqs.(\ref{B}) and (\ref{C}) with
boundary conditions (\ref{D}) and (\ref{E}), respectively. As is
clear from (\ref{B}) and (\ref{C}), these differential equations
becomes more simple if we choose $u'/u = cte$, with solution $u(x)
= \exp(cx)$. To avoid divergency in $x\to \infty$ region ( note
that $x<0$ is not in our physical region, i.e. semi-infinite
laboratory), we must take $c$ negative, $c=-g$, which restricts us
to
\begin{equation}\label{F}
u(x) =e^{-gx}.
\end{equation}

\section{The Klein-Gordon equation}
Since $u(x)$ in eq.(\ref{A}) does not depends on $t,y$, and $z$,
we seek a solution for the Klein-Gordon equation (\ref{B}), with
$u(x)$ defined in (\ref{F}), whose functional form is
$\psiKG(x,y,z,t) = \exp(-iEt + ip_2 y + i p_3 z) \psiKG(x)$. Then
$\psiKG(x)$ satisfies
\begin{equation} \label{G}
\left[ E^2 + \oder{^2}{x^2} - 2g\oder{}{x} - \left(p_2^2 + p_3^2
\right) e^{2gx} - m^2 \right] \psiKG(x) = 0.
\end{equation}
Defining $\phi(x)$ through
\begin{equation} \label{H}
\psiKG(x):= e^{gx}\phi(x),
\end{equation}
we get
\begin{equation} \label{I}
\oder{^2\phi(x)}{x^2}+\left[ \left(\lambda^2 - 1 \right)g^2 -
p^2e^{2gx} \right] \phi(x) = 0,
\end{equation}
in which
\begin{align} \label{J}
p^2:= p_2^2 + p_3^2, \nonumber \\ \lambda^2 :={E^2 - m^2 \over
g^2}.
\end{align}
Changing variable from $x$ to $X=(p/g)\e{gx}$(which is possible
only if $p \neq 0$ ), eq.(\ref{I}) reduces to
\begin{equation} \label{K}
X^2 \oder{^2\phi(X)}{X^2} + X \oder{\phi(X)}{X} - \left(X^2 + 1
-\lambda^2 \right) \phi(X) = 0,
\end{equation}
which is the modified Bessel equation with solutions $K_{\sqrt{1 -
\lambda^2}}(X)$ and $ I_{\sqrt{1-\lambda^2}}(X)$. But the
wavefunctions must satisfy $ \lim_{x\to \infty} \psiKG(x)= 0$,
which restricts us to consider the modified Bessel function
$K_{\nu}(X)$ ($\nu = \sqrt{1 - \lambda^2}$) as solution. Now the
wavefunction of a spin-$0$ particle in a semi-infinite lab must
satisfy (\ref{D}). But $K_{\nu}(X)$ becomes zero only when $\nu$
is pure imaginary, \cite{12}, which means that we have solutions
only when $\lambda^2
> 1$, or we have a ground state with energy:
\begin{equation} \label{M}
E^2 \geq E_{{\rm min}}^2=m^2c^4 + g^2c^2 \hbar^2.
\end{equation}
The wavefunctions are
\begin{equation} \label{N}
\psiKG(x) = e^{gx}K_{i \sqrt{\lambda^2 -1}}({p\over g}e^{gx}),
\end{equation}
and the energy eigenvalues can be obtained by the following
equation
\begin{equation} \label{S}
K_{i \sqrt{\lambda^2 - 1}}({p\over g})=0.
\end{equation}
If $p=0$, the previous change of variable is forbidden. But in
this case, the differential equation (\ref{I}) becomes
\begin{equation} \label{R}
\oder{^2\phi(x)}{x^2}+\left(\lambda^2 -1 \right)g^2 \phi(x) =0.
\end{equation}
For $\lambda^2<1$, the solutions of (\ref{R}) are $\e{\pm\alpha
x}, (\alpha^2=g^2(1- \lambda^2))$, which no combinations of them
can be found so that $\psiKG ( x \to \infty ) \to 0$ and $
\psiKG(0)=0$. For $\lambda^2>1$, the solutions are $\e{\pm ikx},
(k^2=g^2(\lambda^2 -1))$, which again there exist no combinations
to satisfy the desired boundary conditions. So the solution of the
Klein-Gordon equation exists only when the transverse momentum of
particle is different from zero:
\begin{equation} \label{T}
p \neq 0.
\end{equation}
\section{The Dirac equation}
Like the previous section, we again take $\psiD(x,y,z,t)=\exp(-i E
t + i p_2 y + i p_3 z) \psiD(x)$ in eq.(\ref{C}) with $u(x)$
defined in (\ref{F}). The result is :
\begin{equation} \label{X}
\left[-i E \gamma^0 - g \gamma^1 + \gamma^1 \oder{}{x} + i e^{gx}
\left(p_2 \gamma^2 + p_3 \gamma^3 \right) - m \right] \psiD(x) =
0.
\end{equation}
Defining $\tilde\psi_{\rm D}$ through
\begin{equation} \label{Y}
\psiD(x) = e^{gx} \tilde\psi_{\rm D}(x),
\end{equation}
then it satisfies
\begin{equation} \label{Z}
\left[e^{-gx} \gamma^1 \oder{}{x} - i E e^{-g x} \gamma^0 + i
\left(p_2 \gamma^2 + p_3 \gamma^3 \right) - m e^{-g x} \right]
\tilde\psi_{\rm D}(x) = 0.
\end{equation}
Considering the following two operators:
$$ O_1 = e^{-g x} \left(\gamma^1 \oder{}{x} - i E \gamma^0 - m
\right) \gamma^1 \gamma^0 = e^{-g x} \left(\gamma^0 \oder{}{x} - i
E \gamma^1 - m \gamma^1 \gamma^0 \right), $$
\begin{equation}\label{AB}
 O_2 = i \left(p_2 \gamma^2 + p_3 \gamma^3
\right) \gamma^1 \gamma^0,
\end{equation}
it can be easily seen that the eq.(\ref{Z}) is reduced to
following equation for $\phiD(x) = - \gamma^0 \gamma^1
\tilde\psi_{\rm D}(x)$ :
\begin{equation} \label{AC}
\left(O_1 + O_2 \right) \phiD(x) = 0.
\end{equation}
But it can be seen that:
\begin{equation} \label{AD}
\left[O_1 , O_2 \right] = 0,
\end{equation}
so $O_1$ and $O_2$ have the same eigenspinors. Noting that
$\gamma^0 = \left(\begin{matrix}-i & 0 \\ 0 &
i\end{matrix}\right)$ and $\gamma^k = \left(\begin{matrix}0 &
\sigma^k \\ \sigma^k & 0\end{matrix}\right)$, the eigenvalues of
$O_2$ are found to be $\pm ip$, with $p$ defined in (\ref{J}). Let
us focus on $ip$ eigenvalue, which is two-fold degenerate with
eigenspinors :
\begin{equation}
\psi_1 = \left( \begin{matrix} i (p_2 - p)/p_3 \\ 1 \\
0 \\ 0 \end{matrix} \right) \ \ , \ \  \psi_2= \left(\begin
{matrix} 0
\\ 0 \\ i (p_2 + p)/p_3 \\ 1
\end{matrix} \right).
\end{equation}
One can choose $\phi_D$ as $\phi'_1 \psi_1 + \phi_2 \psi_2$, with
arbitrary functions $\phi'_1(x)$ and $\phi_2(x)$, which is
therefore the eigenspinor of $O_2$ with eigenvalue $ip$, i.e. $O_2
\phi_D = i p \phi_D$. If we write $\phi_D = \phi'_1 \psi_1 +
\phi_2 \psi_2 $ as
\begin{equation} \label{AE}
\left(\begin{matrix} \phi_1 \\ \phi'_1 \\ \phi'_2 \\ \phi_2
\end{matrix} \right),
\end{equation}
then $\phi_1$ and $\phi'_2$ are related to $\phi'_1$ and $\phi_2$
as following:
\begin{align} \label{AF}
\phi_1 =\frac{i \left(p_2 - p \right)}{p_3} \phi'_1 ,\\
\phi'_2 = \frac{i \left(p_2 + p \right)}{p_3} \phi_2.
\end{align}
Now $\phi_1$ and $\phi_2$ (and from them $\phi'_1$ and $\phi'_2$ )
are obtained by imposing $\phi_D$ satisfies (\ref{AC}), which
results the following coupled-differential equations:
\begin{align} \label{AG}
\oder{\phi_1}{x} = p e^{g x} \phi_1 - \left(E + m \right) \phi_2,
\\ \oder{\phi_2}{x} = \left(E - m \right) \phi_1 - p e^{g x}
\phi_2.
\end{align}
If one differentiates these two equations, they become decoupled,
and if then changes variable from $x$ to $X=(2p/g){\rm exp}(gx)$
(which is again possible if $p \neq 0$ ), the resulting equation
for $\phi_1$ is:
\begin{equation} \label{AI}
X^2 \oder{^2 \phi_1(X)}{X^2} + X \oder{\phi_1(X)}{X} + \left(-
\frac{1}{2} X - \frac{1}{4} X^2 + \lambda^2 \right) \phi_1(X) = 0,
\end{equation}
where $\lambda $ is defined in (\ref{J}). Defining
$\tilde{\phi}_1$ through
\begin{equation} \label{AJ}
\phi_1 = X^{-1/2} \tilde{\phi}_1 ,
\end{equation}
eq.(\ref{AI}) becomes
\begin{equation} \label{AM}
\oder{^2 \tilde{\phi}_1}{X^2} + \left[- \frac{1}{4} -
\frac{1/2}{X} + \frac{\left(1/4 + \lambda^2 \right)}{X^2} \right]
\tilde{\phi}_1(X) = 0.
\end{equation}
The above equation is Whittaker differential equation with
solution:
\begin{equation} \label{AN}
\tilde{\phi}_1(X) = e^{-X/2} X^{i \lambda + 1/2} \left[c_1 M
\left(1 + i \lambda , 1 + 2 i \lambda , X \right) + c_2 U \left(1
+ i \lambda , 1 + 2 i \lambda , X \right) \right],
\end{equation}
where $M(a,c,x)$ and $U(a,c,x)$ are confluent hypergeometric
functions. Since the asymptotic behavior of $M(a,c,x)$ is
$e^{x}/x^{c-a}$, so $c_1 = 0$. Therefore $\phi_1(X)$ is equal to
\begin{equation} \label{AX}
\phi_1(X) = e^{- X/2} X^{i \lambda} U \left(1 + i \lambda , 1 + 2
i \lambda , X \right).
\end{equation}
$\phi_2(X)$ can be obtained by eq.(\ref{AG}), with result:
$$\phi_2(X) = \frac{g}{m + E} e^{-X/2} X^{i \lambda } [ (1 -
\frac{i \lambda}{X} ) U (1 + i \lambda , 1 + 2 i \lambda , X )$$
\begin{equation} \label{AY} + (1 + i \lambda
) U (2 + i \lambda , 2 + 2 i \lambda , X ) ],
\end{equation}
in which we use the equality:
$$
 \oder{}{x} U (a,c,x )
= - a U (a+1,c+1,x).
$$
The boundary condition (\ref{E}) implies the following boundary
condition on $\phi_1$ and $\phi_2$ :
\begin{equation} \label{AZ}
(\phi_1 + \phi_2 )\mid_{x=0}=0.
\end{equation}
This relation comes from the fact that $\psiD(0) = \tilde
{\psi}_{\rm D}(0) = \gamma^1 \gamma^0 \phiD(0) $. The above
equation can be used to specify the energy eigenvalues. If $p=0$,
the change of variable $x \to X=(2p/g)e^{gx}$ is not applicable,
and the problem must be solved again. We first note that for $p_2
= p_3 = 0$, the operator $O_2$ is equal to zero. So eq.(\ref{AC})
is reduced to $O_1 \phiD(x) = 0$, with result $\e{\pm i p_x x}$
for components of spinor $\phiD$. Therefore $\psiD$ in
eq.(\ref{Y}) becomes proportional to $e^{gx}$ which does not
fulfil the condition $\psiD(x\to\infty)\to 0 $. So like the
Klein-Gordon equation, the Dirac equation in the space-time
considered here, has no solution with $p = 0$ and the transverse
momentum is always different from zero:
$$ p \neq 0. $$
\section{Comparing the spectrums}
The allowed energy eigenvalues of spin-$0$ particles are obtained
by eq.(\ref{S}):
\begin{equation} \label{BA}
K_{i \sqrt{\lambda^2 - 1}}({p \over g}) = 0,
\end{equation}
while for spin-$1/2$ particles, it can be calculated from
eq.(\ref{AZ}) :
\begin{equation} \label{BB}
\phi_1({2p \over g}) + \phi_2({2p \over g}) = 0,
\end{equation}
where $\phi_1$ and $\phi_2$ are given by eqs.(\ref{AX}) and
(\ref{AY}), respectively. Unfortunately, none of these equations
can be solved analytically and only in the first case,
eq.(\ref{BA}), the approximate solutions can be obtained for large
$\lambda$ values.

Obtaining the roots of eqs.(\ref{BA}) and (\ref{BB}) numerically,
for a fixed value of $p/g$ (or $p/(g \hbar)$ in ordinary units),
shows the differences of the spectrums of \lD and \lKG  of Dirac
and Klein-Gordon particles with same mass (see Table 1, as a
specific example). It can also be seen that for any values of $p
/(g\hbar)$, \lKG  is always greater than one (as predicted by
eq.(\ref{M})), but \lD can be less than one (see Table 2).

\begin{table}[here]
\setlength{\tabcolsep}{.3pc} \caption{The first ten values of
$\lD$ and $\lKG$ for $g\equiv 1({\rm meter}^{-1}), m=m_{\rm
electron},$ and $p/(g\hbar)=50$.}
\begin{center}
\begin{tabular}{|c|cccccccccc|}
\hline $\lD$ &26.07 &26.54&27.01 &27.48 &28.41& 28.88 &29.34 &29.79 &30.23 &30.63 \\
\hline $\lKG$ & 26.05 & 26.52& 26.99 & 27.46 & 28.39 & 28.84 & 29.30 & 29.74 & 30.15 & 30.49  \\
\hline
\end{tabular}
\end{center}
\end{table}

\begin{table}[here]
\setlength{\tabcolsep}{.3pc} \caption{The first ten values of
$\lD$ and $\lKG$ for $g\equiv 1({\rm meter}^{-1}), m=m_{\rm
electron},$ and $p/(g\hbar)=0.001$.}
\begin{center}
\begin{tabular}{|c|cccccccccc|}
\hline $\lD$ &0.46 &0.87&1.25&1.61&1.97&2.31&2.66&2.99&3.33&3.66 \\
\hline $\lKG$ &1.08&1.28&1.54&1.83&2.14&2.45&2.76&3.08&3.40&3.71 \\
\hline
\end{tabular}
\end{center}
\end{table}

In $p/(g\hbar)>>1$ region, we can use the asymptotic form of the
zeros of the function $K_{i \alpha}(s)$, to obtain an approximate
values of spectrum of Klein-Gordon equation. It is known that the
relation between $\alpha$ and the zeros of $K_{i \alpha}(s)$ is
\cite{12}
\begin{equation} \label{BC}
\alpha_n = s + \beta_n s^{1/3} + O(s^{-1}),
\end{equation}
in which $\beta_n$ is the $n$-th zero of the Airy function
$\AiryAi(-2^{1/3} \beta)$. In our case (eq.(\ref{BA})), $\alpha_n
= \sqrt{\lambda^2 - 1}$ in which $\lambda_n^2 = (E_n^2 - m^2)/g^2$
(eq.(\ref{J})) and $s = p/g$. So the energy eigenvalues for large
$p/(g\hbar)$ is:
\begin{equation} \label{BD}
E_n^2 = m^2 c^4 + g^2 c^2 \hbar^2 \{ 1 + \left[ {p\over {g \hbar}}
+ \beta_n ({p\over {g \hbar}})^{1/3} \right]^2 + O({g \hbar\over
p}) \}.
\end{equation}
\section{Classical characteristics of the metric}
In this section we are going to investigate the physical and
geometrical properties of the metric:
\begin{equation} \label{BE}
\ds^2 = -\dt^2 + \dx^2 + e^{-2gx} \left (\dy^2 + \dz^2 \right).
\end{equation}
Firstly we obtain the properties of matter corresponds to this
metric, and secondly the geodesics of the above metric. The latter
is important because we have considered semi-infinite laboratory
with a potential barrier at $x=0$. This situation (considering
$x=0$ as the floor of the laboratory) is physical if the classical
particles, in this background metric, strike the barrier, i.e. the
classical trajectories have a falling behavior in -$x$ (downward)
direction.
\subsection{Equation of state}
The non-vanishing
Christofell symbols of the metric (\ref{BE}) are:
\begin{align} \label{BF}
\Gamma^1{}_{22}=\Gamma^1{}_{33}=g e^{-2gx},\nonumber \\
\Gamma^2{}_{12}=\Gamma^2{}_{21}=\Gamma^3{}_{13}=\Gamma^3{}_{31}=-
g.
\end{align}
Therefore the components of Ricci tensor, which are different from
zero, are $R_{11}= 2g^2$ and $R_{22}=R_{33}= 2g^2 e^{-2gx}$, and
the scalar curvature of the metric is:
\begin{equation} \label{BI}
R = g_{\mu \nu} R^{\mu \nu}= 6 g^2.
\end{equation}
Considering the Einstein field equation $R_{\mu \nu} -(1/2) g_{\mu
\nu} R = 8 \pi G T_{\mu \nu}$, the energy-momentum tensor becomes:
\begin{equation} \label{BJ}
T_{\mu \nu}=\kappa \ \ {\rm diag} \left(3, -1, -e^{-2gx},
-e^{-2gx} \right),
\end{equation}
in which:
\begin{equation} \label{BK}
\kappa = \frac{g^2}{8 \pi G}.
\end{equation}
Now consider the energy-momentum tensor of a perfect fluid:
\begin{equation} \label{BL}
T_{\mu \nu} = \left(p + \rho \right) U_\mu U_\nu + p g_{\mu \nu},
\end{equation}
where $U^\mu = (\gamma , \gamma {\bf v})$ is the fluid
four-velocity, and $p$ and $\rho$ are pressure and energy density,
respectively. Comparing eqs.(\ref{BJ}) and (\ref{BL}), shows
$U^\mu = (1,{\bf 0})$, or $U_\mu U_\nu = \delta_{\mu 0}
\delta_{\nu 0}$ (i.e. the fluid is at rest). But $g_{\mu \nu}=$
diag $(-1,1,e^{-2gx},e^{-2gx})$, so $T_{00}=\rho=3\kappa$ and
$T_{ij}=p g_{ij}=-\kappa g_{ij}$. Therefore the matter which is
the origin of the metric (\ref{BE}), is a perfect fluid which is
at rest, has a negative pressure $p=-\kappa$, and its equation of
state is:
\begin{equation} \label{BM}
p = -\frac{\rho}{3}.
\end{equation}
\subsection{The geodesics} The equation of geodesics is:
\begin{equation} \label{BN}
\oder{^2 x^\mu}{s^2} + \Gamma^\mu_{\nu \sigma} \oder{x^\nu}{s}
\oder{x^\sigma}{s}= 0.
\end{equation}
Using the Christofell symbols (\ref{BF}), we arrive at:
\begin{equation} \label{BX}
\oder{^2t}{s^2} = 0,
\end{equation}
\begin{equation} \label{BY}
\oder{^2x}{s^2} + g e^{-2gx} \left[\left(\oder{y}{s} \right)^2 +
\left(\oder{z}{s} \right)^2 \right] = 0,
\end{equation}
\begin{equation} \label{BZ}
\oder{^2y}{s^2} - 2 g \oder{x}{s} \oder{y}{s} = 0,
\end{equation}
\begin{equation} \label{BT}
\oder{^2z}{s^2} - 2 g \oder{x}{s} \oder{z}{s} = 0.
\end{equation}
Equation (\ref{BX}) results ( $c$ is the speed of light):
\begin{equation} \label{BV}
s = c t.
\end{equation}
Differentiating (\ref{BE}) with respect to $s$, and using
eqs.(\ref{BY}) and (\ref{BV}), we find:
\begin{equation} \label{BW}
\frac{{\rm d}u}{2 - u^2} = - g {\rm d}s,
\end{equation}
where $u:={\rm d}x/{\rm d}s$. Noting that $v_x = {\rm d}x/{\rm d}t
< c$ or $u < 1$, integrating of the above equation results:
\begin{equation} \label{CA}
u = \frac{1}{c} v_x = \sqrt{2} \frac{k e^{-\alpha t} - 1}{k
e^{-\alpha t }+ 1},
\end{equation}
where $\alpha=2\sqrt{2}gc$ and $k$ is determined by the initial
condition:
\begin{equation} \label{CB}
k = \frac{c + v_{0x}}{c - v_{0x}},
\end{equation}
where for $v_{0x}>0$, $k$ is greater than one. Once more
integration of (\ref{CA}), specifies $x(t)$ as following:
\begin{equation} \label{CC}
x(t) = x_0 - \frac{1}{2g} \left[\ln \left(e^{\alpha t} + k \right)
\left(k e^{-\alpha t} + 1 \right) - 2 \ln \left(1 + k \right)
\right].
\end{equation}
At $t=t^\ast=(\ln k)/\alpha $, $v_x$ becomes zero and the particle
begins to fall down, i.e. $v_x(t>t^\ast ) < 0$. It can be also
seen that at $t=2t^\ast$, $x(2t^\ast)=x_0$ and $v_x(2t^\ast
)=-v_{0x}$. Therefore everything shows that we have a falling
particle in -$x$ direction.

Equations (\ref{BZ}) and (\ref{BT}) can be written as :
\begin{align} \label{CD}
\oder{v}{s} = 2 g u v, \nonumber  \\ \oder{w}{s} = 2 g u w,
\end{align}
where $v={\rm d}y/{\rm d}s$ and $w={\rm d}z/{\rm d}s$. Dividing
these two equations, results ${\rm d}v/v={\rm d}w/w$, or $v = k_1
w$, from which:
\begin{equation} \label{CE}
y = k_1 z + k_2,
\end{equation}
where $k_1$ and $k_2$ are some constants. Inserting (\ref{CA}) in
(\ref{CD}), gives $v_y$ and $v_z$ as following:
\begin{equation} \label{CF}
v_i = \frac{ (1 + k)^2 v_{0i}}{(e^{\alpha t} + k)(k e^{-\alpha t}
+ 1)},
\end{equation}
with $ i = y , z$. So $v_\perp = 0$ if $v_{0\perp} = 0$.


\begin{thebibliography}{9}
\bibitem{1}  L. D. Landau and E. M. Lifshitz;
              \textit{Quantum Mechanics~--~Non-relativistic Theory}
              (Pergamon Press, 1977).
\bibitem{2}   C. Y. Cardall \& G. M. Fuller; Phys. Rev. {\bf D55}
              (1997) 7960.
\bibitem{3}   M. Alimohammadi \& A. Shariati; Int. J. Mod. Phys.
              {\bf A15} (2000) 4099.
\bibitem{4}   S. Chandrasekhar; Proc. R. Soc. Lond. {\bf A349}
              (1976) 571.
\bibitem{5}   S. K. Chakrabarti, Proc. R. Soc. Lond. {\bf A391}
              (1984) 27.
\bibitem{6}   N. Fornengo, C. Giunti, C. W. Kim, \& J. Song; Phys.
Rev. {\bf D56} (1997) 1895.
\bibitem{7}   M. Alimohammadi \& A. Shariati; Eur. Phys. J. {\bf C21}
(2001) 193.
\bibitem{8}   R. Colella, A. W. Overhauser, \& S. A. Werner;
Phys. Rev. Lett. {\bf 34} (1975) 1472.
\bibitem{9}  V. V. Nesvizhevsky et al; Nature {\bf 415} (2002) 297.
\bibitem{10} K. Varju \& L. H. Ryder; Phys. Lett. {\bf A250} (1998)
263.
\bibitem{11} M. Khorrami, M. Alimohammadi, \& A. Shariati; Annals
of Phys. {\bf 304} (2003) 91 (gr-qc/0210095).
\bibitem{12} W. Magnus, F. Oberhettinger, \& R. P. Soni;\textit{ Formulas
and theorem for the special functions of mathematical physics}
(Springer-Verlag, 1966).
\end{thebibliography}
\end{document}